\documentclass[prb,aps,twocolumn,showpacs,amsmath,amssymb]{revtex4-1}
\usepackage{float}
\usepackage{amsmath}
\usepackage[dvips,final]{graphicx}
\usepackage{epsfig}
\usepackage{color}
\begin{document}

\title{Interface magnetic moments enhancement of FePt-L1$_0$/MgO(001): an $ab$ $initio$ study}

\author{R. Cuadrado}
\affiliation{Department of Physics, University of York, York YO10 5DD,
             United Kingdom}

\author{R. W. Chantrell}
\affiliation{Department of Physics, University of York, York YO10 5DD,
             United Kingdom}

\date{\today}

\begin{abstract}
The interface between FePt--L1$_0$ and MgO(001) alloys has been studied
using density functional calculations. Because the stacking of the
face-centered tetragonal L1$_0$ phase is formed by alternating Fe and Pt
planes, both the Fe- and Pt-terminated contact layers were studied.
Furthermore, due to the large mismatch between the in-plane lattice constants
of both systems, we have chosen some common $a$ values for both alloys in order
to explore in detail the adsorption energy, the electronic structure and the
interface magnetism. The adsorption energy has been calculated by subtracting
the energy of clean FePt and MgO alloys from the total energy. The preferred
adsorbed geometric sites for Fe/Pt atoms are when they lie on $top$ of the O
species, having a smaller adsorption energy for the remaining positions. We found
that expanding the MgO lattice enhances the magnetic moment of the Fe species but
the Pt moments remain almost constant.
\end{abstract}


\maketitle

\section{Introduction}\label{introduction-sec}
The face-centered tetragonal~({\it fct}) L1$_0$ phase of the
3{\it d}--5{\it d} binary based alloys such as FePt have recently been
the subject of much attention because of their potential applications
for the fabrication of ultrahigh density data recording media~\cite{weller_para}.
These alloys present high values of the magnetocrystalline anisotropy~(MAE)
constants (7$\times$10$^7$~ergs/cm$^3$)~\cite{Pirama,ivanov} along the
$c$--axis, a preferred orientation direction~(easy axis). These high values of
the anisotropy are necessary to overcome the superparamagnetic limit~\cite{Yan}
in order to avoid the loss of recorded information. There are several methods
to achieve the desired ferromagnetic structures such as the alternating
monatomic layer deposition of Fe and Pt~\cite{Shima} or alternatively the
room temperature deposition of disordered {\it fcc} FePt on an underlayer
followed by annealing at around 600 $^\circ$C to induce a phase transformation
from {\it fcc} to {\it fct} L1$_0$ stacking. Extensive studies such as the
effect of the alloying composition and growing FePt films on various underlayers,
for instance MgO,~\cite{Chen} PtMn,~\cite{Chiang} or Si,~\cite{Wu} among others,
have been carried out to optimize the microstructure and the magnetic properties
as well as decreasing the processing temperature~\cite{Xu}. For practical purposes
there are still some challenges, for example the ordered FePt grains with
perpendicular (001) crystallographic orientation have to be magnetically decoupled
from each other for which some materials are added in the fabrication process~(see
 Ref.~\cite{Peng} and references therein). The use of  MgO as an underlayer has
some practical drawbacks such as its elevated costs, but the MgO single crystalline
substrate promotes the out-of-plane anisotropy in contrast to other substrates
which tend to promote in-plane anisotropy.~\cite{Perumal} Additionally, because
the MgO lattice parameter is larger than that of FePt, it can easily promote the
$c_{FePt}$ to remain perpendicular during the growing process, thus reducing the
in-plane variants.

From a theoretical point of view the transition metal~(TM) binary
compounds have been extensively studied in the past decade in their
bulk phases, slabs,~\cite{Chepulskii2012} gas phases,~\cite{Cuadrado,
antoniak,rollman} and interfaces~\cite{Zhu1}, however only a few
theoretical studies based on FePt--L1$_0$/MgO(001) interfaces have been
carried out~\cite{Zhu1,Zhu2}. In this work we scan different configurations
for this interface in a systematic investigation of the potential adsorption
geometries, namely, the Fe-/Pt-termination, different atomic adsorption sites
and the influence within the electronic structure of changes in the MgO(001)
lattice constant, $a_{MgO}$.

The paper is structured as follows: The employed theoretical tools are
explained briefly in section~\ref{tools-sec}. In section~\ref{structure-sub-sec}
we will summarize the final geometric structures as well as the related
adsorption energies. The electronic study is presented in the
section~\ref{PDOS-sub-sec} and the magnetic behaviour in~\ref{MMs-sec}.
The conclusions and future work are in section~\ref{conclusions-sec}.

\section{Theoretical Methods}\label{tools-sec}
\begin{figure}[thb]
 \includegraphics[scale=0.47]{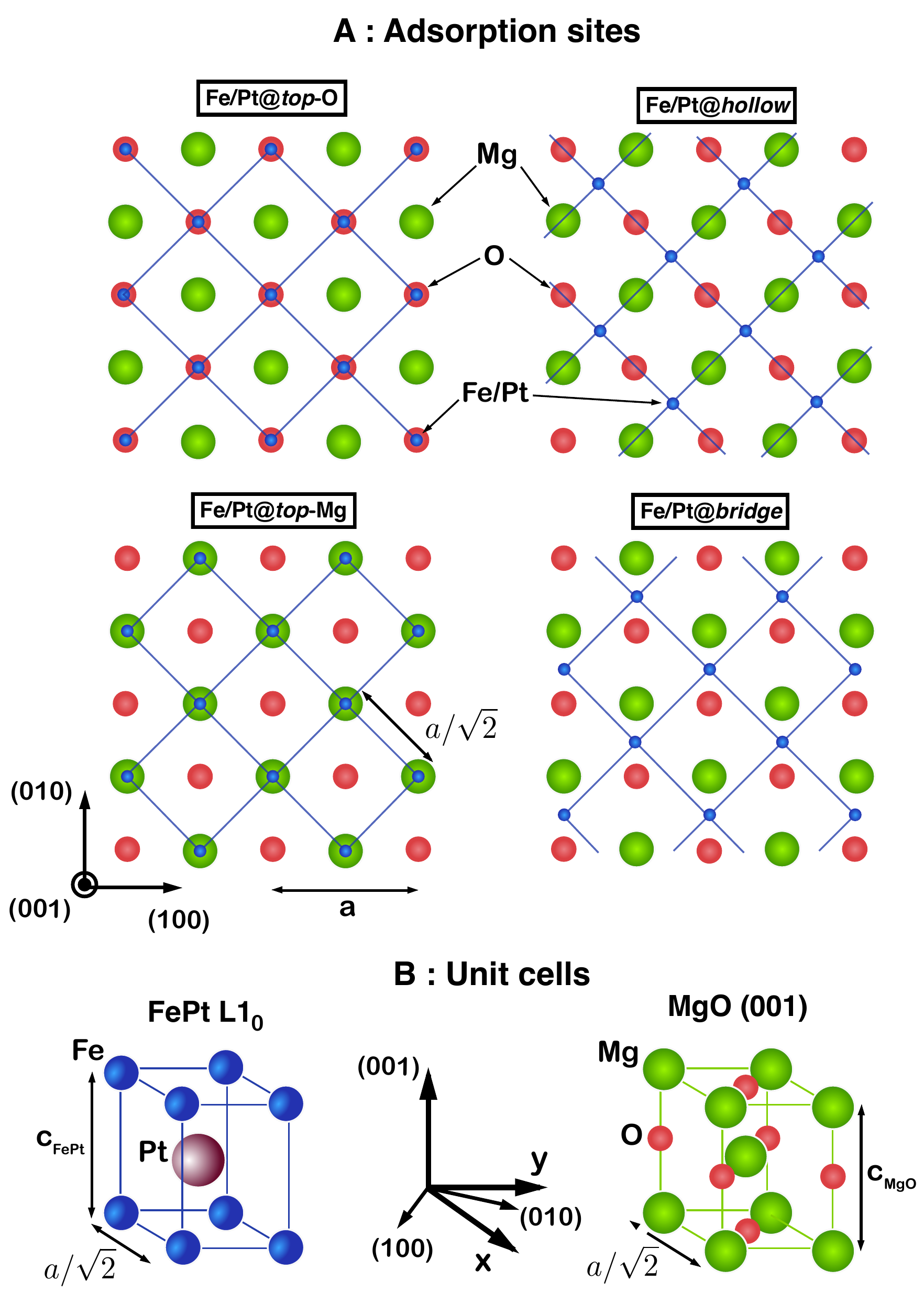}
 \caption{(Color online) ({\bf A}) Schematic top view of the four initial
          adsorption sites for FePt--L1$_0$ onto MgO(001) alloy. Only the atoms
          belonging to the interfaces are shown, i.e., big green and small red
          spheres for the MgO contact layer and blue spheres on whether Fe- or
          Pt-terminations is shown. ({\bf B}) FePt--L1$_0$ and MgO unit cells,
          left and right, respectively. The in-plane lattice constant, $a$, as
          well as some representative orientations and values have been also
          depicted in the figure.}
 \label{fig-geom}
\end{figure}
We have undertaken geometrical, electronic and magnetic structure
calculations of the FePt--L1$_0$/MgO(001) interface by means of DFT
using the SIESTA~\cite{siesta} code. To describe the core electrons we
have used fully separable Kleinmann-Bylander~\cite{kb} and norm-conserving
pseudopotentials~(PP) of the Troulliers-Martins~\cite{tm} type. Our DFT
based calculations have been performed within the generalized gradient
approximation~(GGA) for the exchange correlation~(XC) potential following
the Perdew, Burke, and Ernzerhof~(PBE) version~\cite{pbe}. To address the
description of magnetic systems, pseudocore~(pc) corrections were used to
include in the XC terms not only the valence charge density but also the
core charge~\cite{cc}. In order to ease the convergence of three center
integrals with the size of the real space grid, $\rho^c(r)$ is replaced by
a pseudo-core charge density, $\rho^{pc}(r)$, which equals the real core
charge density beyond a given radius, $r_{pc}$, while close to the nuclei
it becomes a smooth function. The radius $r_{pc}$ should be chosen small
enough to ensure that the overlap region between the valence and the core
charges is fully taken into account. Based on previous studies of the
binary alloys~\cite{LS-paper}, we have chosen for this radius the values
of r$_{pc}$(Fe) = 0.6~Bohrs and r$_{pc}$(Pt) = 1.0~Bohrs, ensuring
that the overlap region between the valence and the core charge is fully
taken into account. As the basis set, we have employed double-zeta
polarized~(DZP) strictly localized numerical atomic orbitals~(AO).
The confinement energy, $E_c$, defined as the energy cost to confine
the wave function within a given radius, was set to 100~meV. The so--called
electronic temperature --kT in the Fermi-Dirac distribution-- was set to
50~meV. In all the cases we ensured convergence of the Brillouin Zone~(BZ)
integration by considering a k--supercell of (16x16), i.e., 256 k-points.
Real space three-center integrals are computed over a three-dimensional
grid with a resolution of 900~Ry, a mesh fine enough to ensure convergence of
the magnetic properties.

The interface system is described by a two-dimensional periodic slab
comprising eight MgO(001) plus eight FePt--L1$_0$ oriented layers.
The L1$_0$ structure stacking is a {\it fct} phase in which there are
alternate planes of Fe and Pt along the (001) direction. Because of this,
there arises the possibility of having two kinds of interfaces between
FePt--L1$_0$ and MgO(001): Fe-terminated~(Fe/MgO) and Pt-terminated~(Pt/MgO).
Furthermore we can see in the Fig.~\ref{fig-geom}A schematically the top
views of four possible initial configurations, first and second rows.
The small red and big green spheres represent the MgO alloy and the blue
represent the first  contact plane of the FePt alloy, either Fe or Pt.
On the other hand, in the Fig.~\ref{fig-geom}B the unit cells are shown
following the same atomic nomenclature as in A but now explicitly the Pt
atom presents a bigger size than the Fe ones. Specifically, depending whether
the Fe/Pt atoms lie on $top$, $hollow$ or at $bridge$ positions, we have
named the configurations as follows: on $top$ of O (Fe/Pt@$top$-O), on
$top$ of Mg (Fe/Pt@$top$-Mg), at $hollow$ (Fe/Pt@$hollow$) and in $bridge$
positions (Fe/Pt@$bridge$). Metastable adsorption structures were obtained
after relaxing the different proposed models to local minima until forces
on atoms were smaller than 0.03 eV/\AA. During the minimization process,
just two layers of each material were allowed to relax leaving the rest of
the atoms in the slab fixed to their bulk positions.

The presence of two kinds of atoms in the FePt-L1$_0$ phases generates a
vertical distortion so that its structure is defined by two quantities,
the in-plane lattice parameter, $a$, and the out-of-plane constant, $c$,
whose bulk experimental values are $a_{FePt}$=3.86~\AA\ and $c/a$=0.98.
The magnesium oxide structure can be described as two inter--penetrating
{\it fcc} lattices displaced by $a/2(111)$ along the body diagonal of the
conventional cube and the bulk experimental value for its lattice parameter
is $a_{MgO}$=4.22\AA. We observe that the in-plane mismatch between
both alloys is $\approx$8.5$\%$. Because of this and in order to scan
more geometrical possibilities we have used the four common in-plane lattice
values: 4.00\AA, 4.05\AA, 4.10\AA\ and 4.30\AA. The last value corresponds
to the optimized lattice constant for the bulk MgO under GGA and it has
been taken into account to address  how the FePt geometry and magnetic
properties change not only with intermediate common $a$ values but also
with an $a_{MgO}$ optimized value. According to the changing $a$, the
distance between planes will vary too, so it was necessary to optimize
the $c$ parameter for each lattice value. The results of both bulk
systems and their corresponding out-of-plane distortions are for MgO:
c/a = 1.14, 1.12, 1.10, 1.00 and for FePt: c/a = 0.92, 0.90, 0.88,
0.75 for a = 4.00\AA, 4.05\AA, 4.10\AA\ and 4.30\AA, respectively.

\section{Results}\label{results-sec}

\subsection{DFT structural relaxations}\label{structure-sub-sec}
\begin{table*}
\begin{tabular}{cccccccccccccc}
       &&&\multicolumn{4}{c}{{\bf Fe-terminated}} &&&\multicolumn{4}{c}{{\bf Pt-terminated}}  &\\ \hline \hline
   Site&a (\AA)&&E$_{ads}$&$z_I(Fe)$&MM$_M$&MM$_{NM}$& &&E$_{ads}$&$z_I(Pt)$&MM$_M$&MM$_{NM}$ \\ \hline
 @$top$-O & {\bf 4.00} && 0.89& 2.23& 3.22& 0.24&&& 0.57 & 2.55 & 3.26& 0.25 \\
          & {\bf 4.05} && 0.93& 2.21& 3.26& 0.25&&& 0.60 & 2.57 & 3.30& 0.26 \\
          & {\bf 4.10} && 0.97& 2.19& 3.29& 0.25&&& 0.62 & 2.56 & 3.34& 0.26 \\
\vspace{0.05cm}
          & {\bf 4.30} && 1.14& 2.16& 3.34& 0.21&&& 0.74 & 2.47 & 3.39& 0.20 \\
 @$top$-Mg& {\bf 4.00} && 0.19& 3.30& 3.26& 0.25&&& 0.22 & 3.16 & 3.27& 0.25 \\
          & {\bf 4.05} && 0.20& 3.47& 3.30& 0.25&&& 0.22 & 3.13 & 3.30& 0.25 \\
          & {\bf 4.10} && 0.20& 3.51& 3.33& 0.25&&& 0.23 & 3.14 & 3.34& 0.26 \\
\vspace{0.05cm}
          & {\bf 4.30} && 0.22& 3.54& 3.37& 0.21&&& 0.29 & 3.07 & 3.40& 0.21 \\
\vspace{0.05cm}
@$hollow$ & {\bf 4.30} && 0.57 & 2.47&3.36 & 0.21&&& 0.45 & 2.68 &3.40 & 0.21 \\
@$bridge$ & {\bf 4.30} && 1.13 & 2.17&3.34 & 0.21&&& 0.70 & 2.51 &3.39 & 0.20 \\ \hline \hline
\end{tabular}
\caption{Adsorption energies, E$_{ads}$, $z$ heigths between the MgO
         contact layer and the first FePt-L1$_0$ plane, $z_I(Fe/Pt)$,
         average magnetic moments~(MM) per atom of the Fe and Pt species of each
         configuration for Fe-/Pt-terminations, columns 3 to 6 and 7 to 9,
         respectively. The first two columns represent the four adsorption
         sites and the common lattice in-plane $a$ values, respectively.
         Energies are in eV, heights in \AA\ and MM in $\mu_B$/at.
         \label{ads-energies}}
\end{table*}
In the table~\ref{ads-energies} we present the results of the interface
adsorption energies E$_{ads}$, the perpendicular bond distance $z_I$
computed as the $z$ difference of the MgO plane and the Fe/Pt one:
$z_I$(A$_i$)=$z_{A_i}$--$z_{B_i}$~[A$_i$=Fe,Pt; B$_i$=Mg,O] and the total
magnetic moment~(MM) per magnetic/non-magnetic species for all the configurations.
Two different contact layers, namely, Fe- and Pt-termination have been taken
into account with a common in-plane lattice constant ranging from 4.00 to
4.30\AA. The adsorption energies were evaluated after subtracting from the
total energy of the FePt+Mg(001) bilayer the energy of the two clean metallic
slabs.

A general tendency of all @$top$ relaxed structures is that as the in-plane
lattice constant expands from 4.00\AA\ to 4.30\AA\ the E$_{ads}$ increases
significantly. It is noticeable that for the Fe-terminated configuration on top
of O~(first four values in the third column of Table~\ref{ads-energies}),
the average adsorption energies are larger by 0.35, 0.87 and 0.74~eV
compared to those of Pt@$top$-O, Fe@$top$-Mg and Pt@$top$-Mg sites,
respectively. Consequently, the bond between the FePt and MgO will be stronger
for the Fe@$top$-O, with decreasing stability for the other cases, with
Fe@$top$-Mg having the lowest stability. In order to scan more possible
accommodation sites for Fe/Pt atoms we also studied Fe/Pt@$hollow$ and
Fe/Pt@$bridge$ adsorption configurations~(see Fig.~\ref{fig-geom}A) considering
only the GGA optimized value of $a_{MgO}$. The E$_{ads}$ values in these cases
are between of those of Fe/Pt@$top$-O/-Mg sites, providing a weak chemical
bonding for FePt. Regarding the Fe/Pt@$bridge$ case and after relaxing the
interfaces,  we found that the strong Fe/Pt-O chemical interaction means that
the Fe/Pt@$bridge$ atoms move closer to the @$top$-O positions. In fact, the
equilibrium position for the Fe/Pt atoms is only $\sim$0.2~\AA\ distant from
their @$top$-O positions.

We also notice, by inspection of the table, that E$_{ads}$ values are quite
similar in both cases. Related to the E$_{ads}$, the bond distance, $z_I$(A$_i$),
will depend on the strength of the bond: the higher the adsorption energies the
smaller $z$ distances and stronger bonding will arise. This implies that the
Fe/Pt planes will be closer to the MgO contact layer resulting in a complex
rearrangement of the charge, changing the final values of the magnetic moments
and hence the magnetic behaviour. Except for the case of Fe@$top$-Mg, where
E$_{ads}$ remains approximately constant, the interlayer bond distance decrease
as $a$ increases. We will discuss in detail in section~\ref{MMs-sec} the behaviour
of the magnetic moments. As a global tendency we can say that the MM$_M$ values
are augmented as $a$ is increased, by an average amount of $\sim$0.13$\mu_B$ and
remain almost constant for MM$_{NM}$ with the special characteristic reduction
of these values for $a$=4.30\AA.

\subsection{Density of states and hybridisation study}\label{PDOS-sub-sec}
\begin{figure*}
 \includegraphics[scale=0.62]{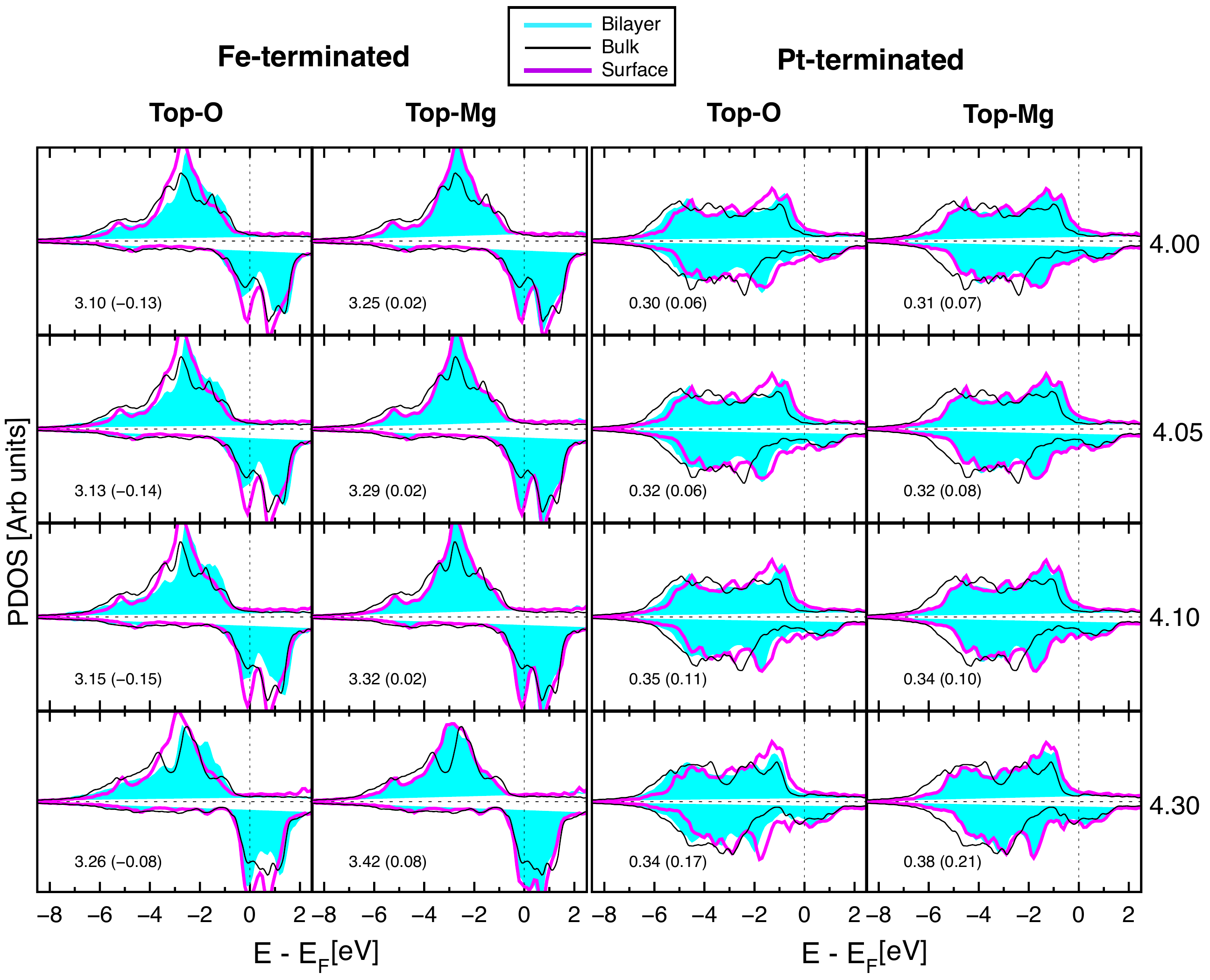}
 \caption{(Color online) Density of states~(DOS) projected on Fe/Pt atoms
          of the FePt--L1$_0$ interfaces~(filled turquoise curve), clean
          surfaces~(thin solid pink line) and the Fe/Pt atoms in their bulk
          phases~(thin solid black line). The Fe-/Pt-terminated configurations
          are presented in the first and second and in the third and fourth
          columns, respectively. For each interface termination @$top$-O/-Mg
          adsorption sites are also shown. The numbers present the
          MM in $\mu_B$/at and in parenthesis their difference with respect to
          the bulk atomic values.}
 \label{pdos-fig}
\end{figure*}

We display in Fig.~\ref{pdos-fig} the evolution of the spin-resolved
density of states~(DOS), projected on the Fe/Pt interface atoms, for Fe- and
Pt- terminated FePt layers. For both cases results are presented for the @$top$-O
and @$top$-Mg adsorption sites. The next FePt--L1$_0$ layers after the contact
layer~(Pt atoms for Fe-termination or Fe for Pt-termination) are not shown here
because changes in the electronic states are not significant beyond the interface
layer. The three different colours in each graph depict the Fe/Pt DOS of the
bilayer~(filled turquoise curve), the Fe/Pt for the clean surface~(thick solid
pink line) and the atomic DOS projection of the atoms in their bulk phase~(thin
solid black line). Inside each graph the MM values appear on the bottom left
corner just for the bilayer configurations together with a positive~(negative)
value within parenthesis that represents the increase~(decrease) of the local
MM compared to the Fe/Pt atomic bulk values.
A first inspection of Fig.~\ref{pdos-fig} shows that the DOS at the interface
layer is significantly affected for contact with both MgO and vacuum. Further
inspection of the DOS for the four rows~(O and Mg adsorption sites) regarding
both the Fe-/Pt-terminations, shows that, when the Fe/Pt atoms lie @$top$-Mg
rather than @$top$-O, the interface Fe/Pt PDOS profile is almost the same as
those of clean surfaces either spin-up or spin-down. Consequently the surface
states of the FePt alloy will not be altered significantly. As a result, when
 Fe/Pt are @$top$-Mg the FePt termination behave similarly to a vacuum
termination irrespective of the contact with MgO. This confirms the significant
difference of $\sim$0.74~eV between the adsorption energies regarding the Mg
or O sites~(see table~\ref{ads-energies}). Then, as we pointed out in the
previous section, the bonding between Fe/Pt and Mg atoms will be weaker than
with the O atoms.

The abrupt termination of the FePt--L1$_0$ alloy in the clean surfaces and in
the interface of the bilayer induces within the $d$ electrons a rearrangement
which is distinct from that in the bulk phase Although the Fe/Pt coordination
will be the same for the whole system, the environment will be modified compared
to that of the bulk, given that now the Mg and O atoms are in the former positions
of the Fe/Pt species. This is noticeable if we inspect the DOS of the Fe-termination
(first two columns). In the bulk case the up-states are located in an energy range
of 5~eV, i. e., from -1 to -6~eV for the @$top$-O/-Mg configurations. Despite the
fact that the DOS of the interfaces and clean surfaces are quite similar
to those of the bulk, their states are located closer to a pronounced peak at
3~eV weakly present in the bulk, leading to a narrowing of both the bilayer and
clean surface {\it d}-bands.

The fact that the spin-up and spin-down black curves have a small shift
to lower energy values, compared to the pink and turquoise curves, in the
last two columns of Fig.~\ref{pdos-fig}~(Pt-termination) can be explained by
inspection of the Mulliken population of the bulk FePt. This shows that the bulk
atomic species are more charged than the other two systems by an average of
$\sim$0.2~e/at.

\begin{figure}
 \includegraphics[scale=0.37]{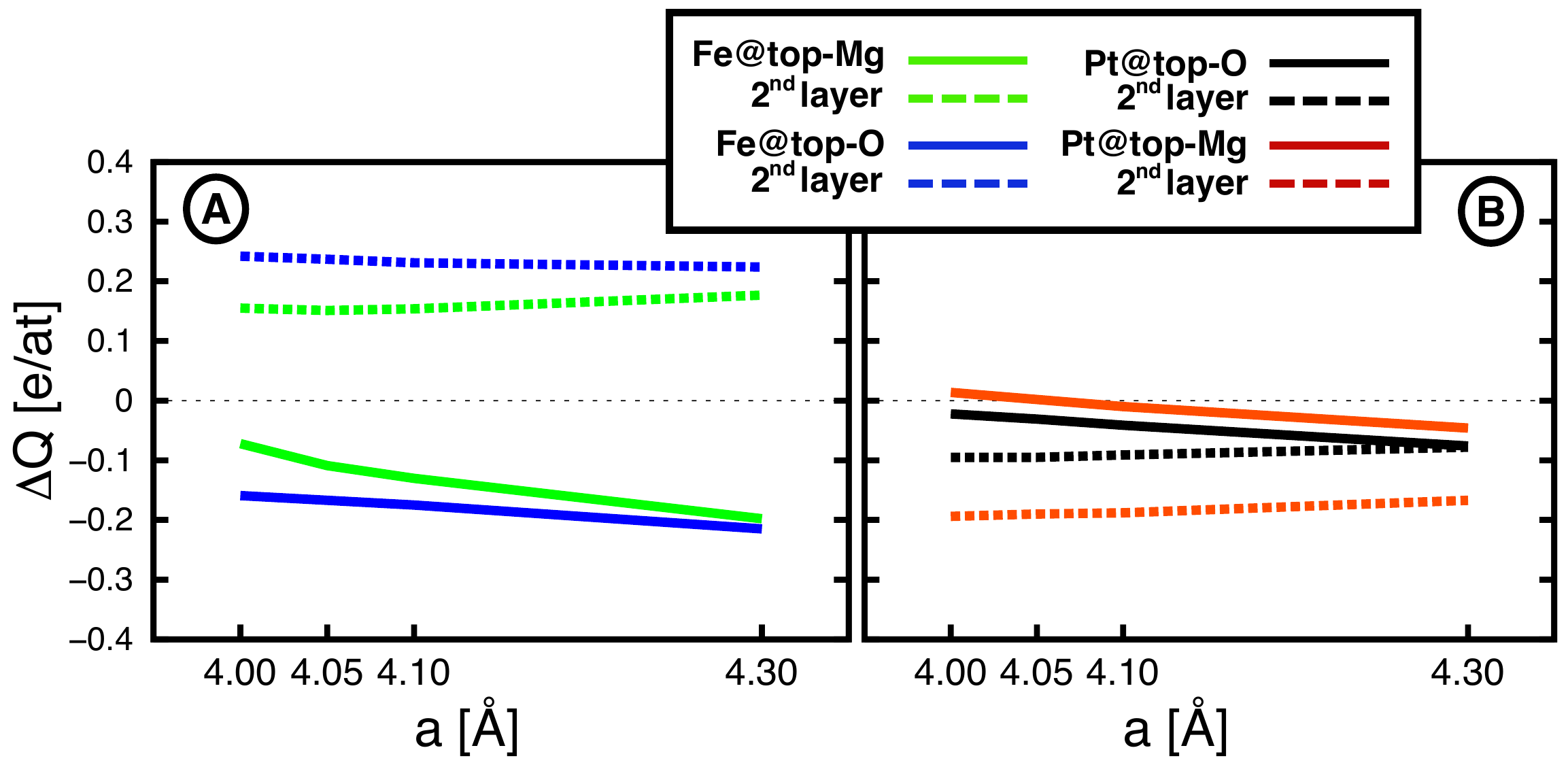}
 \caption{(Color online) Fe and Pt atomic charge difference with respect
          to their bulk phase counterparts,
          $\Delta Q_{Fe/Pt}=q_{Fe/Pt}^{bilayer}-q^{bulk}_{Fe/Pt}$, as a
          function of the in-plane lattice values for the Fe-/Pt-terminated
          configurations, left and right, respectively. The solid coloured
          lines depict the evolution of $\Delta Q_{Fe/Pt}$ interface layer
          @top-O/-Mg adsorption sites and the dashed ones those of the
          nearest inward layer: Pt for Fe-terminations and Fe for
          Pt-termination.}
 \label{q_net-fig}
\end{figure}

The analysis of the data shows that the charge transfer between the
atoms~(orbitals) belonging to an interface is of paramount importance. It
gives us a way to understand the hybridisation between atoms and how this
influences the bond strength and the magnetic behaviour. In figure~\ref{q_net-fig}
it shown that the difference in charge transferred between the magnetic and
non--magnetic species is larger compared to the bulk phases. In figure~\ref{q_net-fig}A,
situating the Fe atoms in the @$top$-O~position (blue solid line) increases the
amount of charge given to the Pt layer~(dashed blue line) and to the first MgO
plane by an average of 0.18~e/at. The same situation occurs when the Fe is
@$top$-Mg~(solid green line) though the effect is smaller. The dispersion ranges
from 0.07~e/at up to almost 0.2~e/at for 4.30\AA. When the Pt contact layer is
@$top$-O~(black solid line) it behaves in a similar way to bulk phase so that
only for the 4.30\AA\ lattice spacing is the amount of excess charge transferred
to the Pt is significant at 0.08~e/at. Similar behavior is shown when comparing the
Fe contact configuration in A. The Fe atoms in this case~(black dashed line)
are responsible for the charge transfer, having almost a constant contribution
along all the $a$ values. Finally, the charge transferred to the Pt atoms
@$top$-Mg~(red solid line) shows very little change with respect to the bulk
phase, increasing only slightly for the large $a$ values. The increase is clear
however for the Fe atoms: again they lose more charge than in the bulk. In
summary, the presence of the MgO changes significantly the FePt-L1$_0$
behavior depending on whether the contact layer is Fe or Pt and also
where the Fe/Pt atoms are located after relaxation. This implies different
kinds of hybridisation between 3$d$ and 5$d$ orbitals of these Fe and Pt atoms
respectively.

In order to link the MM behaviour with the DOS curves we also observe that
in Table~\ref{ads-energies} the MM values of the Fe atoms increase by
0.13$\mu_B/at$ with the in-plane lattice spacing for all the configurations.
This small enhancement of the MM$_M$ can be observed in the DOS curves noting
that the down-state peaks below the Fermi level at $\sim$0.2~eV for $a=$4.00\AA\
move to lie just at the Fermi level for $a=$4.30\AA\ resulting in a deficit of
down-states compared to the up-states which are almost constant with $a$. Finally,
the DOS projected onto the interface plane of the MgO alloy~(not shown here) does
not change significantly, there is only some charge transfer among the
interface resulting in a very small value of the local MM values~(see
Sec.~\ref{MMs-sec}).

\subsection{Interface magnetism: Magnetic Moments}\label{MMs-sec}
\begin{figure}[b]
 \includegraphics[scale=0.36]{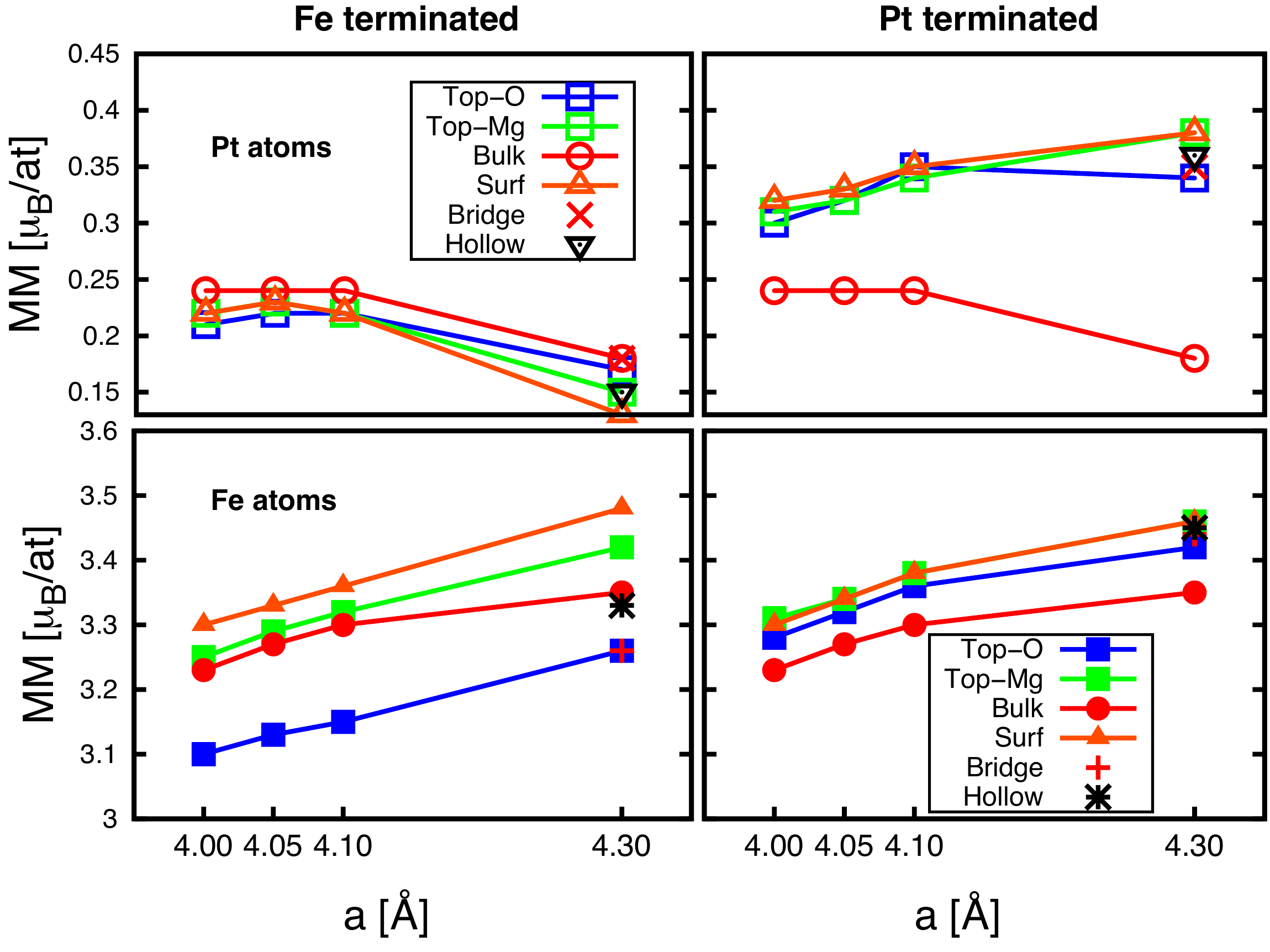}
 \caption{(Color online) Resolved magnetic moments of Pt and Fe atoms,
          first and second row, respectively, for both Fe-~(left) and
          Pt-terminations~(right) as a function of the in-plane lattice
          parameter $a$~(in \AA). For a common type of symbol the empty
          symbols represent the MM of the non magnetic species and the
          filled symbols, those of the Fe atoms. Each kind of color depicts
          @$top$-O adsorption site~(blue squares), @$top$-Mg~(green
          squares), the bulk~(red circles) and the clean surface~(orange
          triangles). The crosses and asterisks show the $bridge$ and
          $hollow$ adsorption sites for 4.30\AA, respectively.}
 \label{fig-MMs}
\end{figure}

We provide in Fig.~\ref{fig-MMs} the magnetic moment~(MM) values
per atom as a function of the in-plane lattice parameter $a$ for all
the configurations studied in this work. The global trend of the MM
values of the Fe atoms for both Fe-/Pt-terminations~(bottom row) is
an enhancement of around 0.13$\mu_B$/at as the lattice increases from
4.00\AA\ to 4.30\AA. Although this behaviour is shared by all the
adsorption positions, i.e., @(top-O, top-Mg,hollow,bridge) there is
just an increase of 0.06$\mu_B$/at in the bulk case~(filled red circles)
when $a$=4.30\AA\ compared to the smaller lattice considered. The values
of the MM of the Pt atoms for both Fe-/Pt-terminations~(first row) are
enhanced as are those of the Fe atoms when $a$ increases up to $a$=4.15\AA.
However the MM decreases~(increases) for the Fe-terminated alloys
(Pt-terminated) for larger $a$. Inspecting the MM values vertically~(common
lattice) either @top-O or @top-Mg, blue and green filled squares, respectively,
we observe that the dispersion between these two adsorption sites is about
0.15$\mu_B$/at. It is evident after inspecting table~\ref{ads-energies}
that the lower distance between the Fe/Pt:MgO layers gives the higher adsorption
energy. This behaviour plays an important role in the net MM/at values
since the charge rearrangement between down and up states of the system
promotes the reduction or the increase of the MM/at depending if the
adsorption site is @top-O or @top-Mg, respectively. Completing
the magnetic study we have added dots to represent the MM values for the
Fe/Pt@($hollow$,$bridge$) sites, black asterisks and red crosses,
respectively. Their MM values do not have significant changes regarding
others with same $a$ value, they are almost the same
as those obtained for the Fe/Pt@$top$-O configurations.

\begin{table}
\begin{tabular}{ccccccccccc}
       &&&\multicolumn{2}{c}{{\bf Fe-terminated}} &&&\multicolumn{2}{c}{{\bf Pt-terminated}}  &\\ \hline \hline
      Site&    a(\AA)  &&MM$_{Mg}$&MM$_O$&&&MM$_{Mg}$& MM$_O$ \\ \hline
 @$top$-O & {\bf 4.00} && -0.11   & 0.03 &&& -0.03   &  0.03  \\
          & {\bf 4.05} && -0.12   & 0.03 &&& -0.03   &  0.03  \\
          & {\bf 4.10} && -0.12   & 0.03 &&& -0.04   &  0.03  \\
\vspace{0.05cm}
          & {\bf 4.30} && -0.10   & 0.04 &&& -0.03   &  0.04  \\
 @$top$-Mg& {\bf 4.00} && -0.00   & 0.02 &&& -0.03   &  0.02  \\
          & {\bf 4.05} && -0.02   & 0.01 &&& -0.03   &  0.02  \\
          & {\bf 4.10} && -0.02   & 0.01 &&& -0.03   &  0.02  \\
\vspace{0.05cm}
          & {\bf 4.30} && -0.01   & 0.01 &&& -0.04   &  0.03  \\
\vspace{0.05cm}
@$hollow$ & {\bf 4.30} && -0.05   & 0.04 &&& -0.02   &  0.04  \\
@$bridge$ & {\bf 4.30} && -0.10   & 0.04 &&& -0.03   &  0.03  \\ \hline \hline
\end{tabular}
\caption{Local magnetic moments~(MM) values of the Mg~(MM$_{Mg}$) and
         O~(MM$_O$) atoms of the MgO contact plane in $\mu_B$. The
         nomenclature as well as the different adsorption sites and
         terminations are the same as used throughout this work.
         \label{MMs-MgO}}
\end{table}

Finally we show in Table~\ref{MMs-MgO} the site resolved MM values
of Mg and O atoms. Note that the Mg species favors the down states while
the O sites exhibit the opposite behavior. The most significant difference
between the distinct configurations occurs @$top$-O adsorption sites
since the MM$_{Mg}$ are almost constant, at a value of -0.12~$\mu_B$
in comparison with the rest. The O species have an average constant
value of $\sim$0.03~$\mu_B$.

\vspace{0.2cm}
\section{Conclusions and future work}\label{conclusions-sec}
We have carried out a first principles study of the FePt--L1$_0$/MgO(001)
interface regarding different in-plane lattice constant values
and the two possible FePt contact planes due to the L1$_0$ stacking.
In addition we scanned the magnetic/electronic properties for different
adsorption sites for a common Fe/Pt plane, namely,
@($top$-O,$top$-Mg,$hollow$,$bridge$). The adsorption energies provide
a way to elucidate the preferential adsorption site and lattice
constant value. The higher values were obtained for Fe@$top$-O
followed by those of the Pt@$top$-O and also when the larger
in-plane lattice constant has been taken into account. The stronger
chemical bonding then occurs for these configurations having a significant
overlap between the $d$-bands and the MgO orbitals. Additionally, we
found that as the E$_{ads}$ decreases(increases) the MM$_M$
values augment(diminish) however the induced MM within the Pt species
is not significantly affected by this variation. The use of GGA as the
XC could overestimate the bonding between these materials giving the
possibility that from an experimental point of view, the bond distance
would be shorter and hence the adsorption energy higher.

A useful compliment to the discussion of atomic hybridisation at the end
of section~\ref{PDOS-sub-sec} would be a study of the up/down population
of the $d$-orbitals. It is however beyond the scope of this work to include
such details and instead we have only explored qualitatively the atomic
spin-resolved population to understand how the spin charge influences
the magnetic behavior. The spin electronic analysis confirms the results of
the DOS calculations~(Fig.~\ref{pdos-fig}). It shows that with increasing
lattice constant, $a$, the amount of up~(down) charge increases~(decreases)
leading to an enhancement of the MM values.

In addition, figure~\ref{fig-MMs} shows that there is a constant difference
of 0.25$\mu_B$/at in the MM values for different adsorption positions~(see
separate blue and green symbols in the Fe-terminated configuration). The
lowest adsorption energy value is shown for the Fe@$top$-O and hence this is
the most stable configuration. This means that the hybridisation between the
MgO and the Fe contact layer is stronger, reducing the MM values and promoting
a bigger charge transfer among atoms. For the Pt-terminated configurations, as
was pointed out in section~\ref{structure-sub-sec}, the adsorption energy is
smaller and the orbital hybridisation is smoother, therefore the MM values are
close to each other for all values of $a$.

It has recently been shown \cite{fefept} that the site-resolved Magnetocrystalline
Anisotropy Energy~(MAE) of FePt is associated with the Fe sites. This is due to the
strong 3d Fe -- 5d Pt hybridisation through which the spin-orbit interaction on the
Pt atoms is transferred to the electronic states at the Fe sites. Consequently, the
complex charge transfer processes at the FePt/MgO interface predicted here might be
expected to be reflected in changes in the FePt MAE at the MgO interface. However,
the charge transfer effects seem rather localized to the FePt/MgO interface, so the
effect might be significant only for ultrathin films. However, this is an interesting
possible effect, which is beyond the scope of the current work but certainly worthy
of further investigation.

\section{Acknowledgments}
The authors are grateful to Dr T.J Klemmer for helpful discussions. Financial support of the EU Seventh Framework Programme under grant
agreement No. 281043, FEMTOSPIN, and Seagate Technology is gratefully
acknowledged.


\begin{thebibliography}{References}
\bibitem{weller_para} D.\ Weller and A.\ Moser, IEEE Trans. Magn., {\bf
    36}, 10 (1999).
\bibitem{Pirama}
  N. Piramanayagam
  J. Appl. Phys. {\bf 102}, 011301 (2007)
\bibitem{ivanov}
   O. A. Ivanov, L. V. Solina, V. A. Demshina and L. M Magat
   Fiz. Metal Metalloyed {\bf 35}, 92 (1973).
\bibitem{Yan}
    M. L. Yan, H. Zeng, N. Powers, and D. J. Sellmyer
    J. Appl. Phys. 91, 8471 (2002)
\bibitem{Shima}
   T. Shima, T. Moriguchi, S. Mitani, and K. Takanashi
   Appl. Phys. Lett. {\bf 80}, 288 (2002)
\bibitem{Chen} 
    J. S. Chen, B. C. Lim, J. F. Hu, Y. K. Kim, B. Liu, and G. M. Chow
    Appl. Phys. Lett. {\bf 90}, 042508 (2007)
\bibitem{Chiang}
    C. C. Chiang, C.-H. Lai, and Y. C. Wu
    Appl. Phys. Lett. {\bf 88}, 152508 (2006)
\bibitem{Wu}  
    Y. C. Wu, L. W. Wang, and C. H. Lai,
    Appl. Phys. Lett. {\bf 91}, 072502 (2007)
\bibitem{Xu}
    Y. F. Xu, J. S. Chen, and J. P. Wang
    Appl. Phys. Lett. {\bf 80}, 3325 (2002).
\bibitem{Peng}
    Yingguo Peng, Jian-Gang Zhu, and David E. Laughlin
    J. Appl. Phys. {\bf 99}, 08F907 (2006).
\bibitem{Perumal}
    A. Perumal, Y. K. Takahashi, T. O. Seki, and K. Hono
    Appl. Phys. Lett. {\bf 92}, 132508 (2008).
\bibitem{Chepulskii2012}
   Roman V. Chepulskii and W. H. Butler
   Appl. Phys. Lett., {\bf 100}, 142405, (2012).
\bibitem{Cuadrado}
    R. Cuadrado, and R. W. Chantrell
    Phys. Rev. B {\bf 86}, 224415 (2012)
\bibitem{antoniak}
   Carolin Antoniak, Markus E. Gruner, Marina Spasova, Anastasia V. Trunova, Florian R\"omer, Anne Warland, Bernhard Krumme, Kai Fauth, Shoheng Sun, Peter Entel, Michael Farle and Heiko Wende
   Nature Communications {\bf 2}, 528 (2011).
\bibitem{rollman} 
   Georg Rollman, Markus E. Gruner, Alfred Hucht, Ralf Meyer, Peter Entel, Murilo L. Tiago and James R. Chelikowsky
   Phys. Rev. Lett. {\bf 99}, 083402 (2007).
\bibitem{Zhu1}
    Wanjiao Zhu, Yaowen Liu, and Chun-Gang Duan
    Appl. Phys. Lett. {\bf 99}, 032508 (2011)
\bibitem{Zhu2}
    Wanjiao Zhu, Hang-Chen Ding, Shi-Jing Gong, Yaowen Liu, and Chun-Gang Duan
    J. Phys.: Condens. Matter. {\bf 25}, 396001 (2013)
\bibitem{siesta}
   J.M. Soler, E. Artacho, J.D. Gale, A. Garc\'ia, J. Junquera, P. Ordej\'on and D. S\'anchez-Portal,
   J. Phys.: Condens. Matter, {\bf 14}, 2745, (2002).
\bibitem{kb}
   L. Kleinman and D. M. Bylander,
   Phys. Rev. Lett., {\bf 48}, 1425, (1982).
\bibitem{tm}
   N. Troullier and J. L. Martins,
   Phys. Rev. B, {\bf 43}, 1993, (1991).
\bibitem{pbe}
   J. P. Perdew, K. Burke and M. Ernzerhof,
   Phys. Rev. Lett., {\bf 77}, 3865, (1996).
\bibitem{cc}
   S.G. Louie, S. Froyen and M.L. Cohen,
   Phys. Rev. B, {\bf 26}, 1738 (1982).
\bibitem{LS-paper}
   R. Cuadrado and J. I. Cerdá
   J. Phys.: Condens. Matter {\bf 24}, 086005 (2012).
\bibitem{fefept} 
   C J Aas, P J Hasnip, R Cuadrado, E M Plotnikova, L Szunyogh, L Udvardi  and R W Chantrell
   Phys Rev B, {\bf 88}, 174409 (2013). 
\end{thebibliography}
\end{document}